\documentclass{article}

\usepackage[utf8]{inputenc} 
\usepackage[T1]{fontenc}    
\usepackage{hyperref}       
\usepackage{url}            
\usepackage{amsfonts}       
\usepackage{graphicx}
\usepackage{amsmath,amssymb,txfonts}

\title{Wavefunction collapse induced by gravity in a relativistic Schr{\'o}dinger-Newton model}

\author{Luis A. Poveda\\
  Departemento de F{\'i}sica Centro Federal de Educa\c{c}\~ao Tecnol{\'o}gica de Minas Gerais\\
  Amazonas $5253$, $30421-169$ Belo Horizonte, MG, Brasil\\
  \texttt{poveda@cefetmg.br}
  \\
Luis Grave de Peralta\\
  Department of Physics and Astronomy\\
  Texas Tech University\\
   Lubbock, TX 79409\\
  \texttt{Luis.Grave-de-Peralta@ttu.edu} 
  \\
  Arquimedes Ruiz-Columbi{\'e}\\
  Wind Energy Program\\
  Texas Tech University\\
   Lubbock, TX $79409$\\
  \texttt{Arquimedes.Ruiz-Columbie@ttu.edu}
}

\begin{document}
\maketitle

\begin{abstract}
A relativistic version of the Schr{\"o}dinger-Newton equation is analyzed within the recently proposed Grave de Peralta approach [L. Grave de Peralta, {\em Results Phys.} {\bf 18} (2020) 103318], which include relativistic effects by a parametrization of the non-relativistic hamiltonian, so as to impose that the average kinetic energy of the system coincide with its relativistic kinetic energy. The reliability of this method is tested for the particle in a box. By applying this method to the Schr{\"o}dinger-Newton equation we shows that the characteristic length of the model [L. Di{\'o}si, {\em Phys. Lett}. {\bf 105A} (1984) 199] goes to zero for a mass of the order of the Planck mass, suggesting a collapse of the wavefuncton, induced by gravity.
\end{abstract}

\section{Introduction}
\label{sec:intro}
The Schr{\"o}dinger-Newton equation (SNE)~\cite{Diosi1984,Moroz1998,Harrison2003,Robertshaw2006,Adler2007,Bahrami2014} is a model which describe the time evolution of a Schr{\"o}dinger quantum field coupled to a Newtonian gravitational field, aimed to elucidate the role of gravity on quantum state reduction~\cite{Karolyhazy1966,Penrose1998,Penrose2014}. For a single-point particle the equation is written as
\begin{equation}
i\hbar \frac{\partial\Psi(\mathbf{x},t)}{\partial t} =\left[-\frac{\hbar^2}{2m}\nabla^2-m\Phi(\mathbf{x},t)\right]\Psi(\mathbf{x},t),
\label{Eq01}
\end{equation}
where $\Phi(\mathbf{x},t)$ is the gravitational potential determined by the mass distribution $m|\Psi(\mathbf{x},t)|^2$. 

Eqs.~(\ref{Eq01}) involves three parameters: $\hbar$, the action quantum constant, $G$, the gravitational constant, and $m$ a free parameter that couples both fields. Therefore, the solutions of the SNE are length scaled by
\begin{equation}
l_D=\frac{\hbar^2}{Gm^3},
\label{Eq03}
\end{equation}
here called the Di{\'o}si length, as it was initially proposed by Di{\'o}si~\cite{Diosi1984}, as a measure of the quantum uncertainty of the stationary soliton-like solutions of the SNE. As observed by Di{\'o}si~\cite{Diosi1984}, for a distance of the order of $l_D$ the gravitational force balances the dispersion of the quantum field $\Psi$ by the action of the $p^2/2m$ operator, leading to a stationary state. The $l_D\sim m^{-3}$ dependence point out to the suppression of the macroscopic quantum behavior due to gravitational collapse for a enough massive particle.

Note that when $m$ is equal to the Planck mass ($m_P=\sqrt{\hbar c/G}$), $l_D$ match the Planck distance ($l_P=\sqrt{\hbar G/c^3}$), and the reduced Compton wavelength ($\lambdabar_C=\hbar/m c$) of the particle. Considering that $\lambdabar_C$ is the cut-off for a relativistic quantum field description and that $l_P$ is a fundamental distance, below which gravity is no longer a classical field, the SNE should break down when the mass approach the Planck mass. 

As suggested elsewhere~\cite{Bahrami2014}, the Eqs.~(\ref{Eq01}) is the Newtonian limit of the semi-classical Einstein equations~\cite{Moller,Rosenfeld1963}. Hence, a relativistic treatment require to deal with the Einstein field equations, couple to a Dirac or Klein-Gordon field. Rather, in the present work we use the Grave de Peralta approach~\cite{Grave2020:196,Grave2020:103318,Grave2020:788,Grave2020:14925,Grave2020:065404} which include relativistic effects to the non-relativistic hamiltonian by introducing a parameter which value is fixed by the condition that the average kinetic energy of the system should coincide with its relativistic kinetic energy. 

The paper is organized as follow. In section~\ref{sec:grave}, the Grave de Peral approach to relativity is briefly summarized; then, the method is apply to well known quantum problem of a particle an infinite well in section~\ref{sec:box}; after that, a relativistic version of the Schr{\"o}dinger-Newton eqaution is given and the relativistic Di{\'o}si length is derived in section~\ref{sec:diosi}; finally, some conclusions are given in section~\ref{sec:concl}.

\section{The Grave de Peralta equation}
\label{sec:grave}
Consider the hamiltonian
\begin{equation}
\hat{H}=\hat{K}^\mathrm{(GP)} + \hat{V},
\end{equation}
where 
\begin{equation}
\hat{K}^\mathrm{(GP)}=-\frac{\hbar^2}{\left(1 + \gamma \right)m}\nabla^2,
\label{Eq04}
\end{equation}
is the Grave de Peralta (GP) kinetic energy operator~\cite{Grave2020:196,Grave2020:103318,Grave2020:788,Grave2020:14925,Grave2020:065404} obtained by first quantization ($p\to \hat{p}\equiv\hbar\nabla$) of the relativistic kinetic energy~\cite{Christodeulides} 
\begin{equation} 
\frac{p^2}{\left(1 + \gamma \right)m}
\label{Eq04a}
\end{equation} 
where $\gamma=\left({1+v^2/c^2}\right)^{-1/2}$ is the Lorentz's factor. 

Note that
\begin{equation}
\hat{K}^\mathrm{(GP)}=\frac{2}{\left(1 + \gamma \right)}\hat{K}^\mathrm{(S)},
\label{Eq04b}
\end{equation}
where $\hat{K}^\mathrm{(S)}$ is the well known Schr{\"o}dinger ($\mathrm{S}$) kinetic energy operator, which result from the quantization of the non-relativistic version of Eq.~(\ref{Eq04a}).

In Eqs.~(\ref{Eq04}) and~~(\ref{Eq04b}), $\gamma$ is not an operator, instead is a parameter which value is chosen by the condition
\begin{equation} 
\langle\hat{K}^\mathrm{(GP)}\rangle=(\gamma-1)mc^2
\label{Eq04c}
\end{equation} 
where $\langle~\cdot~\rangle$ is the average in an appropriated quantum state. 

Substituting Eq.~(\ref{Eq04b}) in~(\ref{Eq04c}) and solving for the parameter $\gamma$ we obtain,
\begin{equation} 
\gamma=\sqrt{1+\frac{2\langle \hat{K}^\mathrm{(S)}\rangle}{mc^2}}
\label{Eq04d}
\end{equation} 
and note that in general, $\gamma$ has a coordinate dependence.

The Grave de Peralta equation (GPE) is the Schr{\"o}dinger-like equation, 
\begin{equation} 
i\hbar \frac{\partial\Psi(\mathbf{x},t)}{\partial t} =\left[-\frac{\hbar^2}{(1+\gamma)m}\nabla^2+V(\mathbf{x}) \right]\Psi(\mathbf{x},t)
\label{Eq05}
\end{equation} 
with $\gamma$ given by Eq.~(\ref{Eq04d}). The corresponding stationary solutions fulfills the {\em eigen}value equation
\begin{equation} 
\left[-\frac{\hbar^2}{(1+\gamma)m}\nabla^2+V(\mathbf{x}) \right]\psi(\mathbf{x})=E\psi(\mathbf{x}),
\label{Eq07}
\end{equation} 

In the next section we show that the GdPE include relativistic corrections in a reliable way, for the elementary system of a particle in a one-dimensional box of size $L$\cite{Grave2020:103318}.

\section{Relativistic particle in a box}
\label{sec:box}
In this case, $V(x)=0$ for $0\leq x\leq L$, and the non-relativistic energy {\em eigen}values are given by~\cite{Griffiths},
\begin{equation}
E_n^\mathrm{(S)}=\frac{\hbar^2 n^2\pi^2}{2mL^2}
\label{Eq08}
\end{equation}
Considering that $\langle \hat{H}^\mathrm{(S)}\rangle=\langle\hat{K}^\mathrm{(S)}\rangle$ inside the box, and taking the average in a Schr{\"o}dinger {\em eigen}state, Eq.~(\ref{Eq04d}) adopt the form,
\begin{equation} 
\gamma=\sqrt{1+\frac{\hbar^2n^2\pi^2}{m^2c^2L^2}}
\end{equation}

\begin{figure}
\centering
        \includegraphics[scale=1, width=10cm]{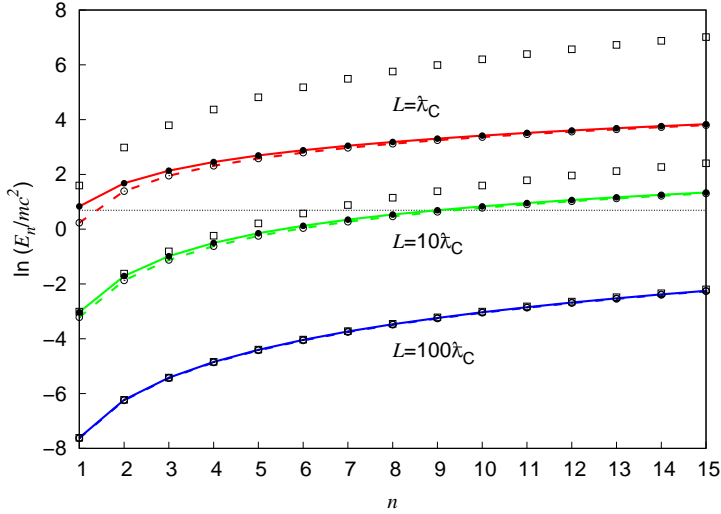}
        \caption{Comparison of the present results (solid dots, solid lines) from Eq.~(\ref{Eq13}) and the reported in ref.~\cite{Alberto1996} (empty dots, dashed lines), for different sizes of the well. In square dots the corresponding non-relativistic energies. The thin line indicates the energy threshold $2mc^2$.}
        \label{fig:1}     
\end{figure}

Finally the energy of the $n$-th level, within the relativistic Grave de Peralta approach will be~\cite{Grave2020:103318}
\begin{equation}
E_n^\mathrm{(GP)}=\frac{\hbar^2 n^2 \pi^2 }{\left[ 1+\sqrt{ 1+\frac{\hbar^2n^2\pi^2}{m^2c^2L^2}}\right]m L^2}
\label{Eq13}
\end{equation}

First, is easy to see that by expanding Eq.~(\ref{Eq13}) in power of $(p/mc)^2$, with $p=\hbar n\pi/L$, up to the second order we obtain,
\begin{equation}
E_n=\frac{\hbar^2 n^2\pi^2 }{2mL^2}-\frac{\hbar^4 n^4\pi^4} {8m^3 c^2 L^4}+\mathcal{O}(L^{-6})
\label{Eq14}
\end{equation}
where the first order coincides with the non-relativistic energy, Eq.~(\ref{Eq08}), and the second order gives the mass-velocity relativistic correction to the kinetic energy.

A comparison of the energies computed with Eq.~(\ref{Eq13}) and those reported in reference~\cite{Alberto1996}, is given in Figure~\ref{fig:1}, for different sizes of the well. Apart from disagreements for lower $n$, the present model reliably predict, with very good agreement, most of the reported data and describe the tendency when the energy increase and the size of the well reduce.

\begin{figure}
\centering
        \includegraphics[scale=1, width=10cm]{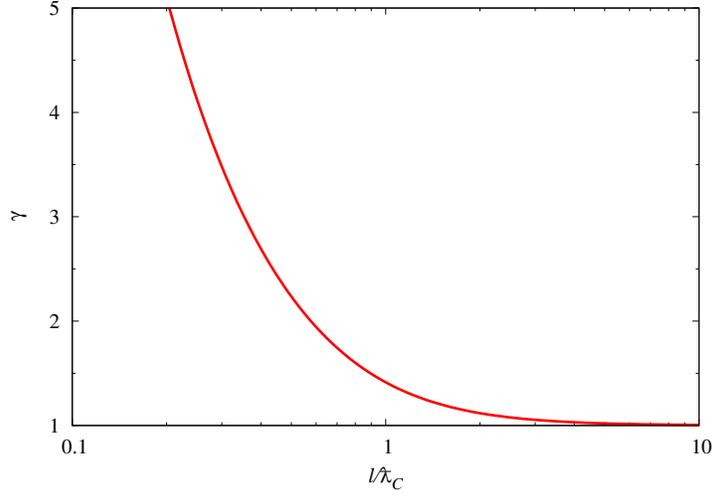}
        \caption{A plot of Eq.~(\ref{Eq15}). See text for details.}
        \label{fig:2}     
\end{figure}

In general por a particle confined in a region of size $l$, the relativistic parameter $\gamma$ of Eq.~(\ref{Eq04d}) can be evaluated as,
\begin{equation}
\gamma=\sqrt{1+\left(\frac{\lambdabar_C}{l}\right)^2}
\label{Eq15}
\end{equation}
where $\lambdabar_C=\hbar/mc$ is the reduced Compton wavelength of the particle. Figure~\ref{fig:2} show the behavior of $\gamma$ as a function of the particle seize in units of $\lambdabar_C$. From the figure, the parameter appreciable deviates from $1$ when the localization region of the particle approach the Compton wavelength. Moreover, when the particle extent to a spatial region several orders larger than its Compton wavelength, the relativistic effects becomes negligible. It is worth noting that for a particle with a mass close to the Planck mass, the relativistic effects start to become relevant when the particle is localized in a region of the order of the Planck length, which coincide with $\lambdabar_C$ for $m=m_P$.  

\section{The relativistic Di{\'o}si length}
\label{sec:diosi}
Now consider the {\em eigen}value equation,
\begin{equation} 
\left[\frac{\hbar^2}{\left(1+\gamma\right)m}\nabla^2- {Gm^2}\int\frac{~|\psi(\mathbf{x'})|^2}{|\mathbf{x}-\mathbf{x'}|}d\mathbf{x'} \right]\psi=E\psi,
\label{Eq17}
\end{equation}
which is the Schr{\"o}dinger-Newton model with the Grave de Peralta modification to include relativistic effects. 

Using Eq.~(\ref{Eq15}) we can qualitatively evaluate the energy of a stationary solution of Eq.~(\ref{Eq17}), for a particle localized in a region of size $l$, in the form,
\begin{equation}
E\approx \frac{\hbar^2}{\left[1+\sqrt{1+\left(\frac{\lambdabar}{l}\right)^2}\right]m l^2}-\frac{Gm^2}{l}.
\label{Eq19}
\end{equation}

This equation is the relativistic version of those previously obtained by Di{\'o}si~\cite{Diosi1984}, namely
\begin{equation}
E\approx \frac{\hbar^2}{2m l^2}-\frac{Gm^2}{l}.
\label{Eq19a}
\end{equation}

By minimizing this expression with respect to $l$, Di{\'o}si arrived to the distance $l_D$ given by Eq.~(\ref{Eq03}), which represent the characteristic length of a stationary solution of the SNE. Indeed, this length give a measure of the size of the quantum field when its expansion, due to the dispersive $p^2/2m$ energy term, is stopped by the gravitational self-force. 

Following Di{\'o}si, it is straightforward to shows that the value of $l$ for which Eq.~(\ref{Eq19}) has a minimum is,
\begin{equation}
l_D^{(r)}=l_D\sqrt{1-\left(\frac{\lambdabar_C}{l_D}\right)^2},
\label{Eq21}
\end{equation}
where the supra-index $r$ in $l_D^{(r)}$ stand for "relativistic". 

\begin{figure}
\centering
        \includegraphics[scale=1, width=10cm]{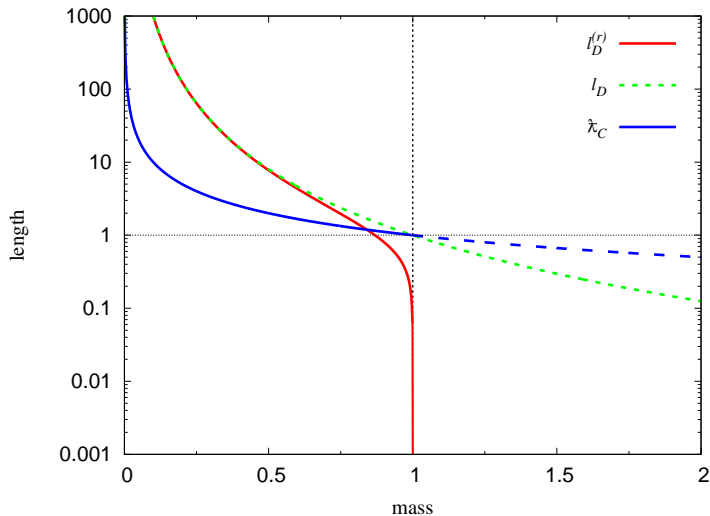}
        \caption{Length scales in units of the Planck length as a function of the mass in units of the Planck mass. See text for details.}
        \label{fig:3}     
\end{figure}

In the same way, $l_D^{(r)}$ represent the length scale for the solutions of Eq.~(\ref{Eq17}), and clearly indicate that this characteristic length goes to zero when the Di{\'o}si length approach the Compton wavelength of the particle. This happens for a mass of the particle equal to the Planck mass, when it is verified that $l_D=\lambdabar_C=l_P$. Figure~\ref{fig:3} shows the behavior of the different length scales in units of the Planck scale, where it is evident that close to the Planck mass the relativistic corrected Di{\'o}si length start to depart from the non-relativistic one and sharply drop to zero, then becoming undefined for larger vales. This sort of collapse behavior observed for the characteristic length, may be given a glimpse of a truly collapse behavior of a quantum field induced by gravity, in a full relativistic approach.  

The Eq.~(\ref{Eq21}) may be written in the following appealing form
\begin{equation}
l_D^{(r)}=l_D\sqrt{1-\left(\frac{m_0}{m_P}\right)^4}
\label{Eq22}
\end{equation}

\section{Conclusions}
\label{sec:concl}

The Grave de Peralta proposal to account for relativistic effects by the Eqs.~(\ref{Eq04d}),~(\ref{Eq05}) and~(\ref{Eq07}), shows a good agreement with the values obtained by integration of the Dirac equation for a particle in a box. Although some appreciable disagreements are observed above the $2m_0c^2$ threshold, our values reproduce with good agreement most of the reported data. The major advantage of this method rely on the possibility of include relativistic corrections by solving just an equation beyond a Schr{\"o}dinger-like equation.

After extend the Schr{\"o}dinger-Newton model to the relativistic domain using the Grave de Peralta approach, we obtained that length scale of the model is bounded from above at a mass of the particle of the order of the Planck mass. This indicates that the coupling between a relativistic quantum field and a classical gravitational field may lead to a collapse of the wavefunction induced by gravity.  

\bibliographystyle{unsrt}  


\begin{thebibliography}{10}

\bibitem{Diosi1984}
L.~Diósi.
\newblock {Gravitation and quantum-mechanical localization of macro-objects}.
\newblock {\em Phys. Lett.}, 105A:199, 1984.

\bibitem{Moroz1998}
I.~M. Moroz, R.~Penrose, and P.~Tod.
\newblock {Spherically-symmetric solutions of the Schrödinger–Newton
  equations}.
\newblock {\em Class. Quant. Grav.}, 15(5):2733, 1998.

\bibitem{Harrison2003}
R.~Harrison, I.~M. Moroz, and K.~P. Tod.
\newblock {Polarisation potentials for positron-molecule scattering processes}.
\newblock {\em Nonlinearity}, 16:101, 2003.

\bibitem{Robertshaw2006}
O.~Robertshaw and P.~Tod.
\newblock {Lie point symmetries and an approximate solution for the
  Schrödinger–Newton equations}.
\newblock {\em Nonlinearity}, 19:1507, 2006.

\bibitem{Adler2007}
S.~L. Adler.
\newblock {}.
\newblock {\em J. Phys. A}, 40:755, 2007.

\bibitem{Bahrami2014}
M.~Bahrami, A.~Großardt, S.~Donadi, and A.~Bassi.
\newblock {The Schrödinger–Newton equation and its foundations}.
\newblock {\em New J. Phys.}, 16:115007, 2014.

\bibitem{Karolyhazy1966}
Karólyházy.
\newblock {Gravitation and quantum mechanics of macroscopic objects}.
\newblock {\em Il Novo Cimento}, 42:390, 1966.

\bibitem{Penrose1998}
R.~Penrose.
\newblock {Quantum computation, entanglement and state reduction}.
\newblock {\em Phil. Trans. R. Soc. Lond.}, 356:1927, 1998.

\bibitem{Penrose2014}
R.~Penrose.
\newblock {On the Gravitization of Quantum Mechanics 1: Quantum State
  Reduction}.
\newblock {\em Foundations of Physics,}, 44:557, 2014.

\bibitem{Moller}
C.~M{\o}ller.
\newblock {\em Les Théories Relativistes de la Gravitation}.
\newblock Colloques Internationaux CNRS 91, 1962.

\bibitem{Rosenfeld1963}
L.~Rosenfeld.
\newblock {On quantization of fields}.
\newblock {\em Nucl. Phys.}, 40:353, 1963.

\bibitem{Grave2020:196}
L.~{Grave de Peralta}.
\newblock {Natural extension of the Schrödinger equation to quasi-relativistic
  speeds}.
\newblock {\em J. Mod. Phys.}, 11:196, 2020.

\bibitem{Grave2020:103318}
L.~{Grave de Peralta}.
\newblock {Quasi-relativistic description of a quantum particle moving through
  one-dimensional piecewise constant potentials}.
\newblock {\em Results Phys.}, 18:103318, 2020.

\bibitem{Grave2020:788}
L.~{Grave de Peralta}.
\newblock {Quasi-relativistic description of Hydrogen-like atoms}.
\newblock {\em J. Mod. Phys.}, 11:788, 2020.

\bibitem{Grave2020:14925}
L.~{Grave de Peralta}.
\newblock {Exact quasi‐relativistic wavefunctions of Hydrogen‐like atoms}.
\newblock {\em Sci. Rep.}, 10:14925, 2020.

\bibitem{Grave2020:065404}
L.~{Grave de Peralta}.
\newblock {Did Schrödinger have other options?}.
\newblock {\em Eur. J. Phys.}, 41:065404, 2020.

\bibitem{Christodeulides}
C.~Christodeulides.
\newblock {\em The Special Theory of Relativity}.
\newblock Springer, New York, 2016.

\bibitem{Griffiths}
D.~J. Griffiths.
\newblock {\em Introduction to Quantum Mechanics, 3rd Ed.}
\newblock Prentice Hall, USA, 2018.

\bibitem{Alberto1996}
P.~Alberto, C.~Fiolhais, and V.~M.~S. Gil.
\newblock {Relativistic particle in a box}.
\newblock {\em Eur. J. Phys.}, 17:19, 1996.

\end{thebibliography}

\end{document}